\documentclass[12pt,a4paper,reqno]{amsart}
\usepackage{graphicx} 
\usepackage{amsmath,amssymb,amsthm,array}
\usepackage{amsfonts, amsmath, amsthm, amssymb} 

\newcommand{\sech}{\mathop{\mathrm{sech}}\nolimits}

\begin{document}
\title[ KdV datum splitting and its conserved quantities] {On connection between the splitting parameters of KdV initial datum and its conservation quantities}

\author{Alexey Samokhin}\vspace{6pt}

\address{Institute of Control Sciences of Russian Academy of Sciences
65 Profsoyuznaya street, Moscow 117997, Russia}\vspace{6pt}

\email{ samohinalexey@gmail.com}\vspace{6pt}

\begin{abstract}
An arbitrary compact-support initial datum for the Korteweg-de Vries equation asymptotically splits into solitons and a radiation tail, moving in opposite direction. We give asimple method to predict the number and amplitudes of resulting solitons and some integral characteristics of the tail using only conservation laws.
  \vspace{1mm}

\noindent\textbf{Keywords:} Korteweg-de Vries equation,  soliton, asymptotic, initial datum splitting.

\noindent\textbf{MSC[2010]:} 35Q53, 35B36.
\end{abstract}

\maketitle

\section{Introduction}

Many physical systems are modeled using equations that admit soliton solutions. Solitons and solitary waves have been observed in numerous situations and often dominate long-time behavior. The behavior of solutions of the KdV and KdV - Burgers equations is a subject of various recent research, \cite{key-1}--\cite{key-4}. The paper is a continuation of the previous research of the author, \cite{key-5} -- \cite{key-10}, that dealt  with inhomogeneity of perturbed media.

In the case of the Korteweg-de Vries equation for an arbitrary compact-support initial datum, it  eventually splits into a number of solitons plus a decaying radiation tai1 moving in opposite direction.  The first numerical evidence for such a behaviour was found by Zabusky and Kruskal \cite{key-11}. First rigorous results  were
proved by Sabat \cite{key-12} and Tanaka \cite{key-13}; for further  history of this problem see \cite{key-14}. The more recent paper  \cite{key-4} gives exact formulas for splitting of so called quenched solitons.

In this paper we give a simple algorithm to predict the number and amplitudes of resulting solitons and some integral characteristics of the tail.
  The main idea is simple enough. Since the resulting solitons and the tail are asymptotically isolated, numerically it makes sense to consider the whole solution as a sum of these solitons and tail (it is also physically reasonable). Then any conserved quantity (infinite number of them) also splits between these summands. The form of every soliton is defined by a single distinct parameter and so do its conserved quantities. This way we obtain a system of equations leading to desired estimations.

 The  the KdV equation considered here is of the form
 \begin{equation}\label{01}
    u_t=2uu_x+u_{xxx}.
    \end{equation}
  The solitary traveling waves (solitons) have a form of peak \[Sol_{a,s}(x,t)=6 a^2\sech^2(a(x+s)+4a^3t)\] and move to the left with velocity $4a^2$ and amplitude $6a^2$. Up to an $s$, a shift of placement on the $x$ axis, the form of a soliton is defined by the parameter $a$. Since $Sol_{a,s}(x,t)\equiv Sol_{-a,s}(x,t)$ we assume $a\geqslant 0$ below.

We use the following initial value - boundary problem  for the KdV-Burgers equation on $x\in \mathbb{R}$:
 \begin{equation}\label{08}
u(x,0) =f(x), \; u(\pm \infty,t) =0,\; u_x(\pm \infty,t) =0.
\end{equation}

We assume that the initial data $u(x,0)$  is bounded and has a compact support.

The asymptotic form (at $t\rightarrow\infty$) of the $N$-soliton solution to this problem is

\[\sum_{i=1}^{N} 6 a_i^2\sech^2(a_ix+p_i+4a_i^3t)+R(x,t),\]

where $R(x,t)$ is a tail and phase shifts are given by the formula

\[p_i=\frac{1}{2}\log\left(\frac{\gamma_i}{2a_i}\cdot\prod_{j=i+1}^N\left(\frac{a_j-a_i}{a_j+a_i}\right)^2\right).\]

Here $\{-a^2_i\}$ is the the discrete specter of the differential operator $-\frac{d^2}{dx^2}-f(x)$ and $\gamma_i$ are the norming constants from the inverse scattering procedure. For an arbitrary $f(x)$ this data is hard to obtain, so estimations, proposed in this paper and  based solely on conserved quantities may be useful.

 For numerical computations we use $ x\in[a,b]$ for appropriately large $a,\;b$ instead of $\mathbb{R}$.

\section{Conservation laws}

\subsection{Soliton's accompanying series}

The first four conserved quantities for KdV are

\begin{eqnarray*}
  I_1(u)) &=& \int_{-\infty}^{+\infty}u(x,t)\,dx \mbox{               --- mass,} \\
  I_2(u) &=& \int_{-\infty}^{+\infty}u^2(x,t)\,dx  \mbox{            --- momentum,} \\
I_3(u) &=& \int_{-\infty}^{+\infty}\left(2u^3(x,t)-3(u_x(x,t))^2\right)\,dx  \mbox{  --- energy,}\\
I_4(u) &=& \int_{-\infty}^{+\infty}\left(5u^4(x,t)-30u(x,t)(u_x(x,t))^2+9(u_{xx}(x,t))^2\right)\,dx,
\end{eqnarray*}
and there are infinite number of them.

There is a simple recurrent procedure to generate $I_k(u)\rightarrow I_{k+1}(u)$ using the bi-hamiltonian structure  of KdV (see \cite{key-14}); note 
that for the KdV of the form $u_t=u_{xxx}+2uu_x$ the hamiltonian operators are $D$ and $(D^3+uD+u_x) $, where $D$ is a total derivative with respect to $x$.

If $u(x,t)$ is a solution of KdV then $\dfrac{\partial}{\partial t} I_k(u)=0$. So if $u(x,t)$ is the solution with the initial value $u(x,0)=f(x)$ then $I_k(u)=I_k(f)$. Thus $I_k(f)$ is conserved in time

In particular, for solitons\newline $u(x,t)=Sol_{a,s}(x,t) =6 a^2\sech^2(a(x+s)+4a^3t)$ we have

\begin{eqnarray}
  I_1(Sol_{a,s}) &=& \int_{-\infty}^{+\infty}6 a^2\sech^2(ax)\,dx = 12a,\\ \nonumber
  I_2(Sol_{a,s}) &=& \int_{-\infty}^{+\infty}(6 a^2\sech^2(ax))^2\,dx = 48 a^3, \\ \nonumber
  I_3(Sol_{a,s}) &=& \frac{1728}{5}a^5,\\ \nonumber
   I_4(Sol_{a,s}) &=& \frac{20736}{7} a^{7} ,\\ \nonumber
   \dots & \dots & \dots \\ \nonumber
    I_l(Sol_{a,s}) &=& K_l a^{2l-1}.
\end{eqnarray}

We obtained the series, common for all KdV solitons, in odd powers of the parameter $a$.

\subsection{Predicting the final pattern of evolution}

If $q(x)=u(x,0)$ is an arbitrary initial
datum with compact support (it is also called a potential), it eventually splits into a
number of solitons plus a decaying radiation tail moving in opposite
direction. A potential without a tail is called reflectionless.

\subsubsection{Reflectionless splitting}

Since after some deliberation $q(x)$ splits (at least numerically) into a disconnected sum of $N$ different-speed solitons,  we get

\begin{eqnarray}\nonumber
 I_1(q)) &=& \int_{-\infty}^{+\infty}q(x)\,dx =\sum_{i=1}^N \int_{-\infty}^{+\infty}Sol_{a_i,s}\,dx=12\sum_{i=1}^N a_i\\ \nonumber
  I_2(q) &=& \int_{-\infty}^{+\infty}q^2(x)\,dx  =48\sum_{i=1}^N a_i^3 \\ \nonumber
I_3(q) &=& \int_{-\infty}^{+\infty}\left(2q^3(x)-3(q_x(x))^2\right)\,dx =\frac{1728}{5}\sum_{i=1}^N a_i^5\\ \nonumber
I_4(q) &=& \int_{-\infty}^{+\infty}\left(5q^4(x)-30q(x)(q_x(x))^2+9q_{xx}(x)^2\right)\,dx =\frac{20736}{7}\sum_{i=1}^N a_i^7\\ \nonumber
\dots&&\dots\label{norefl}
\end{eqnarray}

Thus we obtain the system to obtain  $a_i,\; i=1 \dots N:$
\[K_j\sum_{i=1}^N a_i^{2j+1}=I_j(q), \; j=1 \dots N,\]
where $K_j$ is the constant specific to the $j$-th conserved quantity and $a_1\geqslant a_1> a_2>\dots a_N> 0$ is assumed.

Of course, the above equation hold for all $j=1\dots \infty$, but to find $N$ solitons it suffice to consider only first $N$ equations.

\subsubsection{General case}

If  a reflection is present then the reflected tail eventually disconnects from solitons and \eqref{norefl} holds no more. Instead, we get

\begin{eqnarray}\nonumber
 I_1(q)) &=& 12\sum_{i=1}^N a_i+ \int_{-\infty}^{+\infty}R(x,t)\,dx\\ \nonumber
  I_2(q) &=& 48\sum_{i=1}^N a_i^3+\int_{-\infty}^{+\infty}R^2(x,t)\,dx \\ \nonumber
I_3(q) &=& \frac{1728}{5}\sum_{i=1}^N a_i^5 +\int_{-\infty}^{+\infty}\left(2R^3(x,t)-3(R_x(x,t))^2\right)\,dx \\ \nonumber
\dots&&\dots\label{refl}
\end{eqnarray}

It follows that the discrepancies $I_j(R(x,t))=I_j(q)- K_j\sum_{i=1}^N a_i^{2j+1}$ are also constant.

The first four of  $I_j(R)$ are alternating in sign.
Indeed, at least the initial perturbation mass is carried away by solitons, so $I_1(R)\leqslant 0$; Since momentum of any part of solution is non-negative, it follows that $I_2(R)\geqslant 0$.
The reflected  tail is oscillating around zero value, therefore

$$\int_{-\infty}^{+\infty}\left(2R^3(x,t)\right)\,dx \mbox{ is small while } \int_{-\infty}^{+\infty}\left(-3(R_x(x,t))^2\right)\,dx $$ is negative and comparatively large;
so $I_3(R)\leqslant 0$. Similarly plausible argument can be applied to $I_4(R)$  if  the conservation law is rewritten to equivalent quadratic form
\[5u^4-30uu_x^2+9u_{xx}^2\sim 5u^4+15u^2u_{xx}+9u_{xx}^2=9(u^2+\frac{5+\sqrt{5}}{6}u_{xx})(u^2+\frac{5-\sqrt{5}}{6}u_{xx});\]
the rigorous  proof of alternation in the case of a tail will be published elsewhere.

Hence we obtain the system of necessary conditions

\begin{eqnarray}\nonumber
 I_1(q) &\leqslant& 12\sum_{i=1}^N a_i\\ \nonumber
  I_2(q) &\geqslant& 48\sum_{i=1}^N a_i^3\\ \nonumber
I_3(q) &\leqslant& \frac{1728}{5}\sum_{i=1}^N a_i^5 \\ \nonumber
I_4(q) &\geqslant& \frac{20736}{7}\sum_{i=1}^N a_i^7 \\ \nonumber
\dots&&\dots ;\\ \nonumber
a_i&\geqslant &0.\label{necess}
\end{eqnarray}

The system is very simple and can be effectively used in predicting the number of solitons and their parameters in the resulting splitting. For one, a solution of \eqref{norefl} is a rough approximation to the splitting parameters.

\subsubsection{Number of solitons.}

The system \eqref{necess} defines the admissible domain in $\{x_1,x_2,\dots,x_n\}$, the solitons'parameters space. In order for this domain not to be empty a number of inequalities must hold, for instance $I_1(q) \leqslant 12\sum_{i=1}^N a_i,\;
  I_2(q) \geqslant 48\sum_{i=1}^N a_i^3$. In the case $n=2$ we get $a_1+a_2\geqslant p_1,\; a_1^3+a_2^3\leqslant p_3$ (here $p_k=K^{-1}_jI_j(q)$, $j=2k-1$).

\begin{figure}[h]
 \begin{minipage}{13.2pc}
\includegraphics[width=13.2pc]{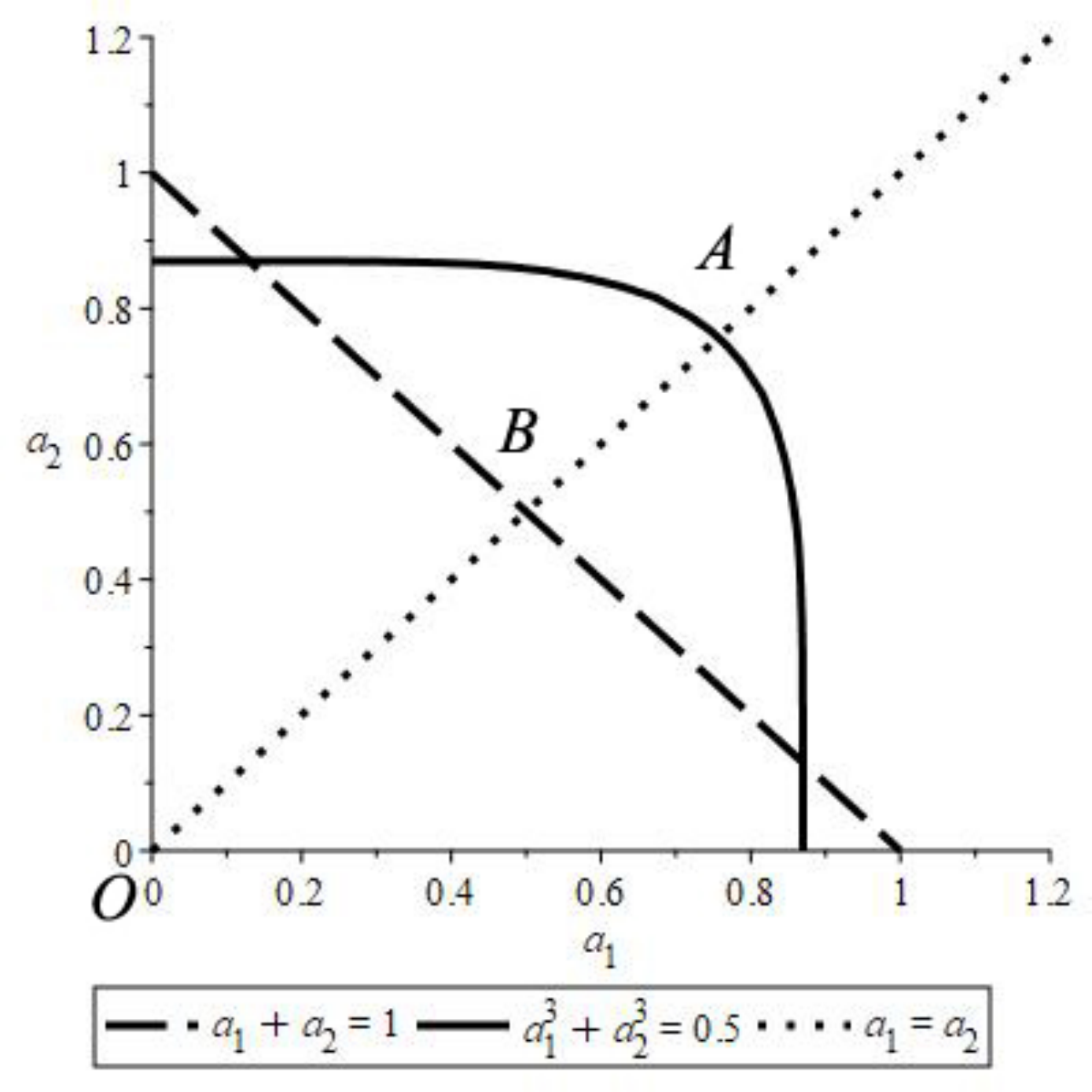}
\end{minipage}
\begin{minipage}{13.2pc}
\includegraphics[width=13.2pc]{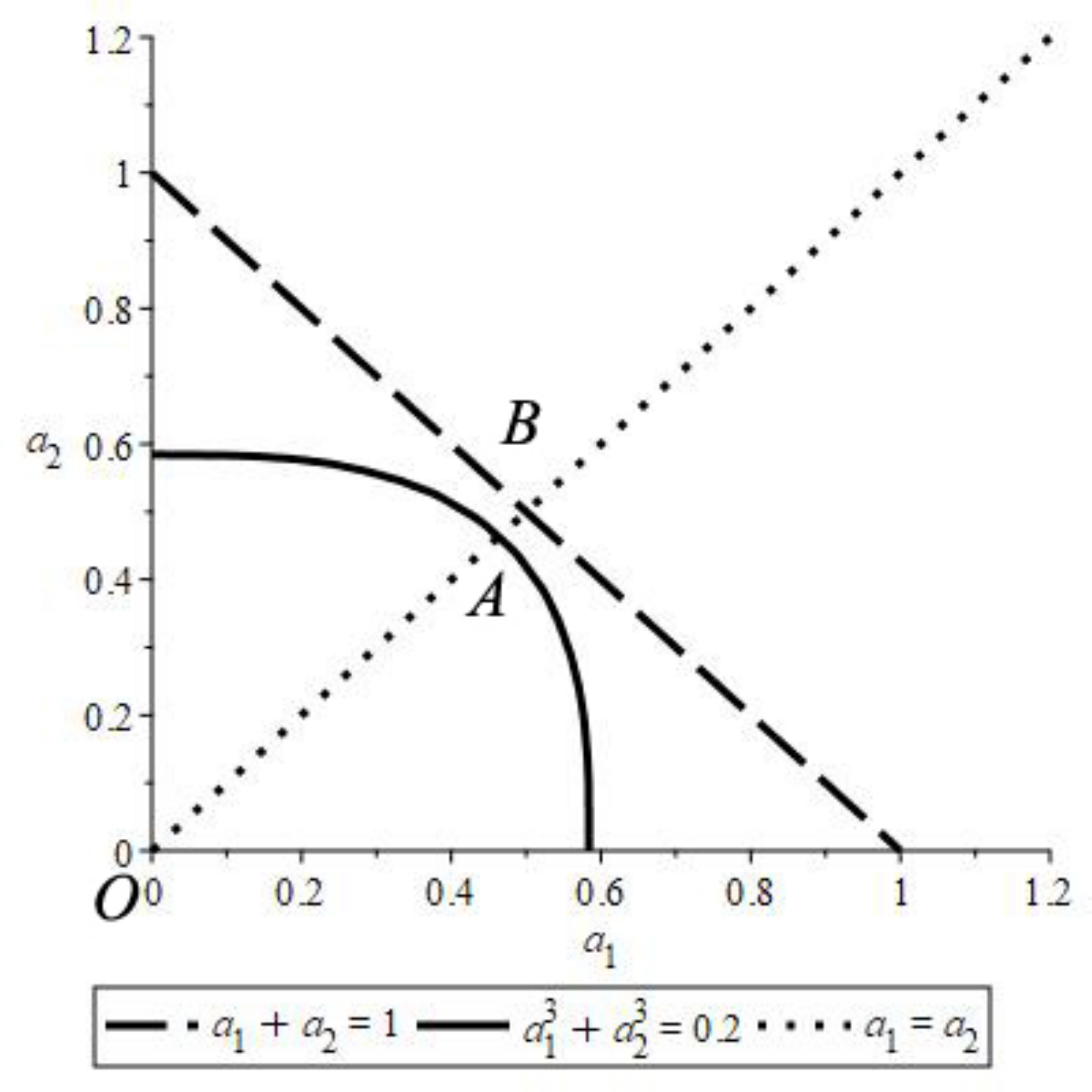}
\end{minipage}
\caption{\textsl{Defining the number of solitons.}\protect\newline \textbf{Left}: $n=2$. \textbf{Right}: $n=3$\label{cond}}
\end{figure}

  The admissible domain would be nonempty if $OA>OB$ as  on the graph \ref{cond} (left),  where $p_1=1,p_2=0.5$. For both points $A$ and $B$ $a_1=a_2$, so $OA=\sqrt{2\left(\frac{p_1}{2}\right)^2}=\sqrt{\frac{1}{2}}$ and $OB=\sqrt{2\left(\frac{p_2}{2}\right)^2}=\sqrt{\frac{1}{8}}$

  But in the case shown on the right part of the figure $p_1=1,p_2=0.2$ the admissible domain is empty. Let's increase the number of solitons to $n$. Then $a_1=a_2=\dots a_n=\frac{1}{n}$ for $A$ and $a^3_1=a^3_2=\dots =a^3_n=\frac{0.2}{n}$.
   Thus if we require $OA^2=n\left(\frac{1}{n^2}\right)\geqslant OB^2=n\left(\sqrt[3]{\frac{0.2}{n}}\right)^2\Rightarrow n^2>5\Rightarrow n=3.$

   For arbitrary $p_1,p_2$ the smallest number of solitons is the integer $n$ such that

   \begin{equation}\label{n}
   n\geqslant \sqrt{\frac{p_1^3}{p_2}}.
   \end{equation}

   For other conserved quantities similar conditions of non-emptiness of the admissible domain lead  to compare $n^{k-1}\vee \frac{p_1^k}{p_2}$. However usually (eg,for all examples below) it suffice to use \eqref{n} to predict the  right number of resulting solitons

\section{Examples}

\subsection{\underline{1-soliton} $q(x)=1+ \cos(x), \mbox{ on } x\in [-\pi,\pi]$\hfill}

In this example $I_1(q)=2\pi,\;I_2(q)=3\pi,\;I_3(q)=7\pi$.

The number of solitons $n>\sqrt{\frac{p_1^3}{p_2}}=\sqrt{\left(\frac{\pi}{6}\right)^3/\frac{\pi}{16}}\thickapprox 0.48\Rightarrow n=1$

\begin{figure}[h]
 \begin{minipage}{13.2pc}
\includegraphics[width=13.2pc]{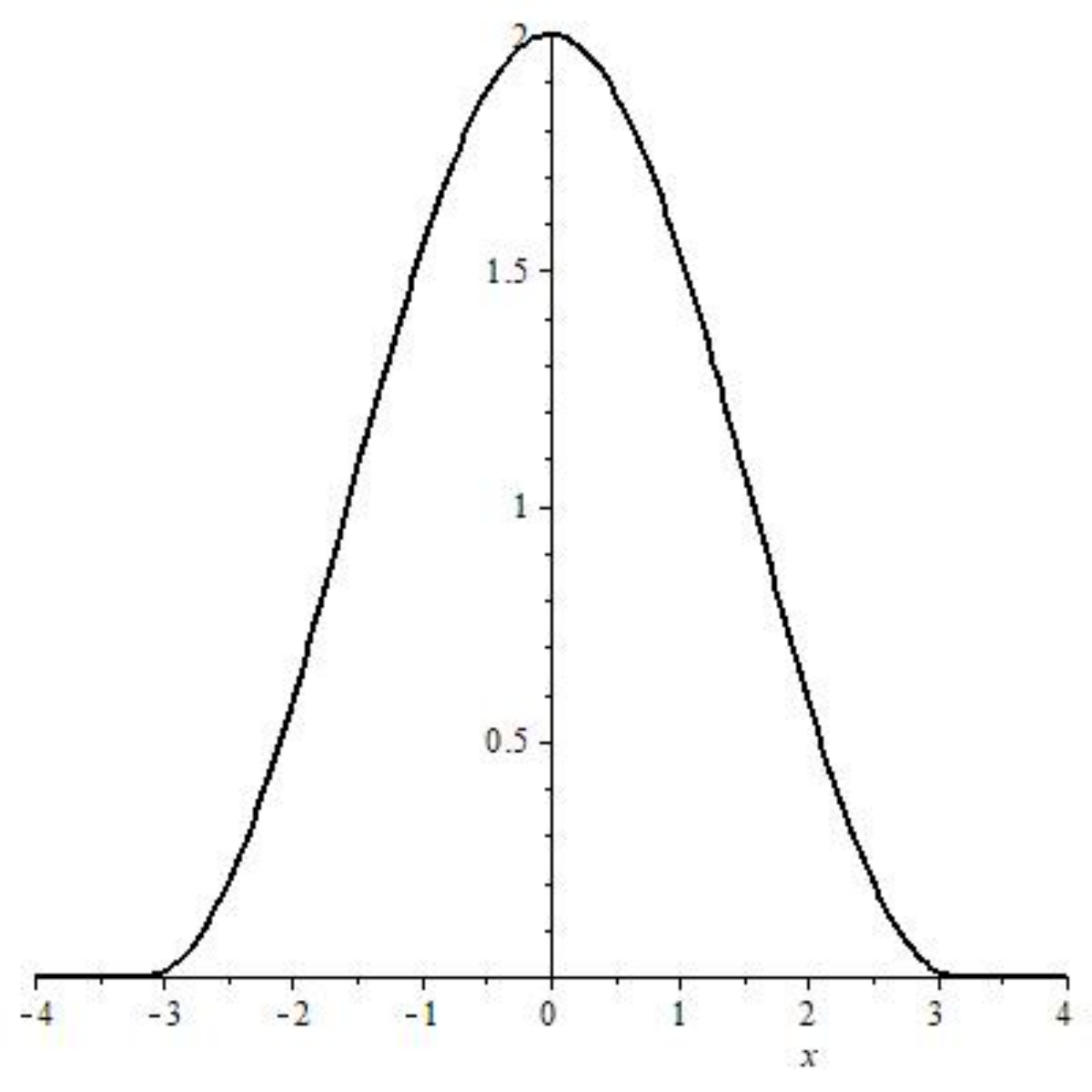}
\end{minipage}
\begin{minipage}{13.2pc}
\includegraphics[width=13.2pc]{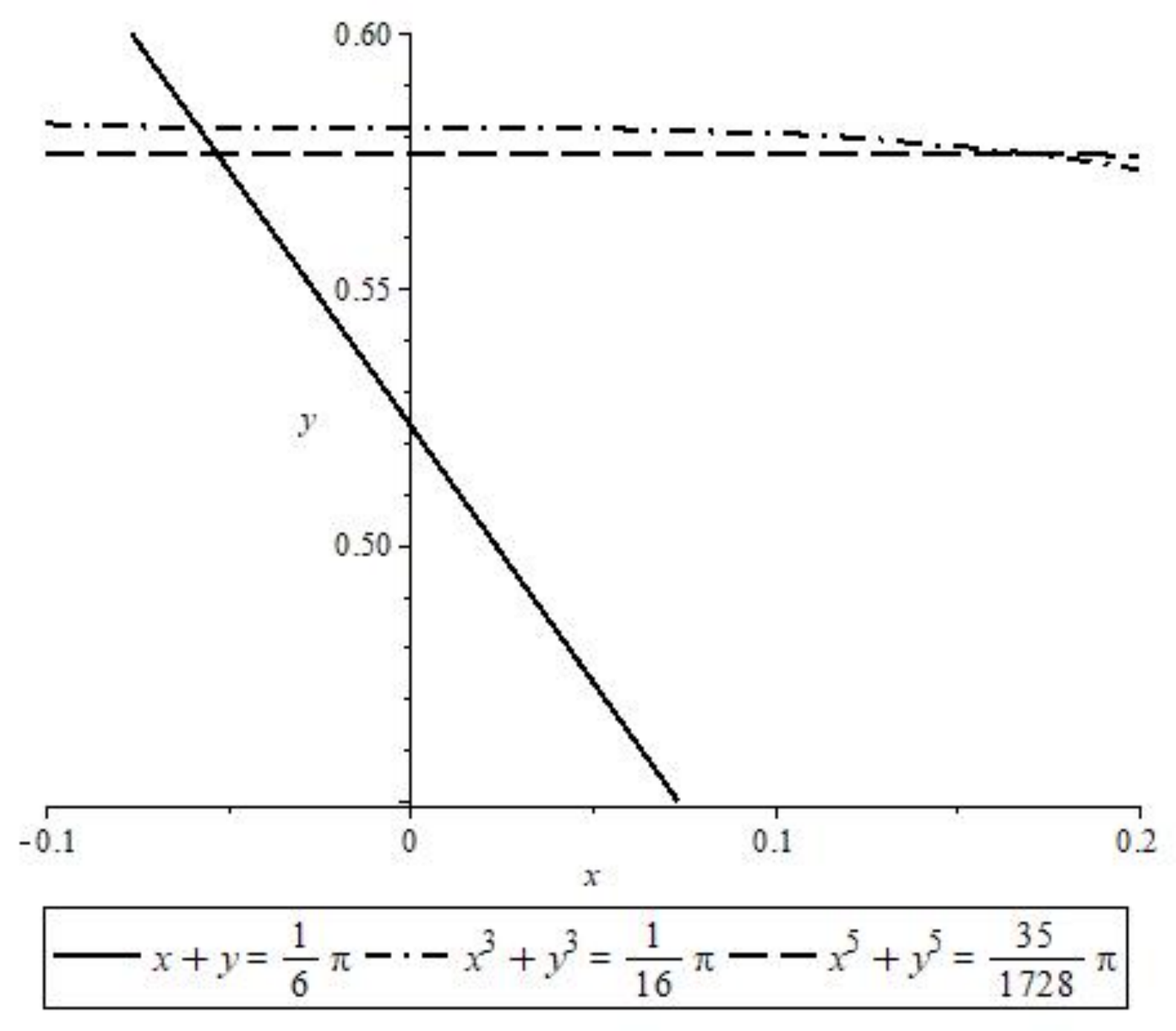}
\end{minipage}
\caption{\textsl{\textbf{Left}: Initial perturbation $q(x)=1+ \cos(x)$; $t=0$.  
\textbf{Right}: Admissible domain is bounded by dash,  dash-dot lines and $y $ axis.}}
\label{28}
\end{figure}

\begin{figure}[h]
 \begin{minipage}{13.2pc}
\includegraphics[width=13.2pc]{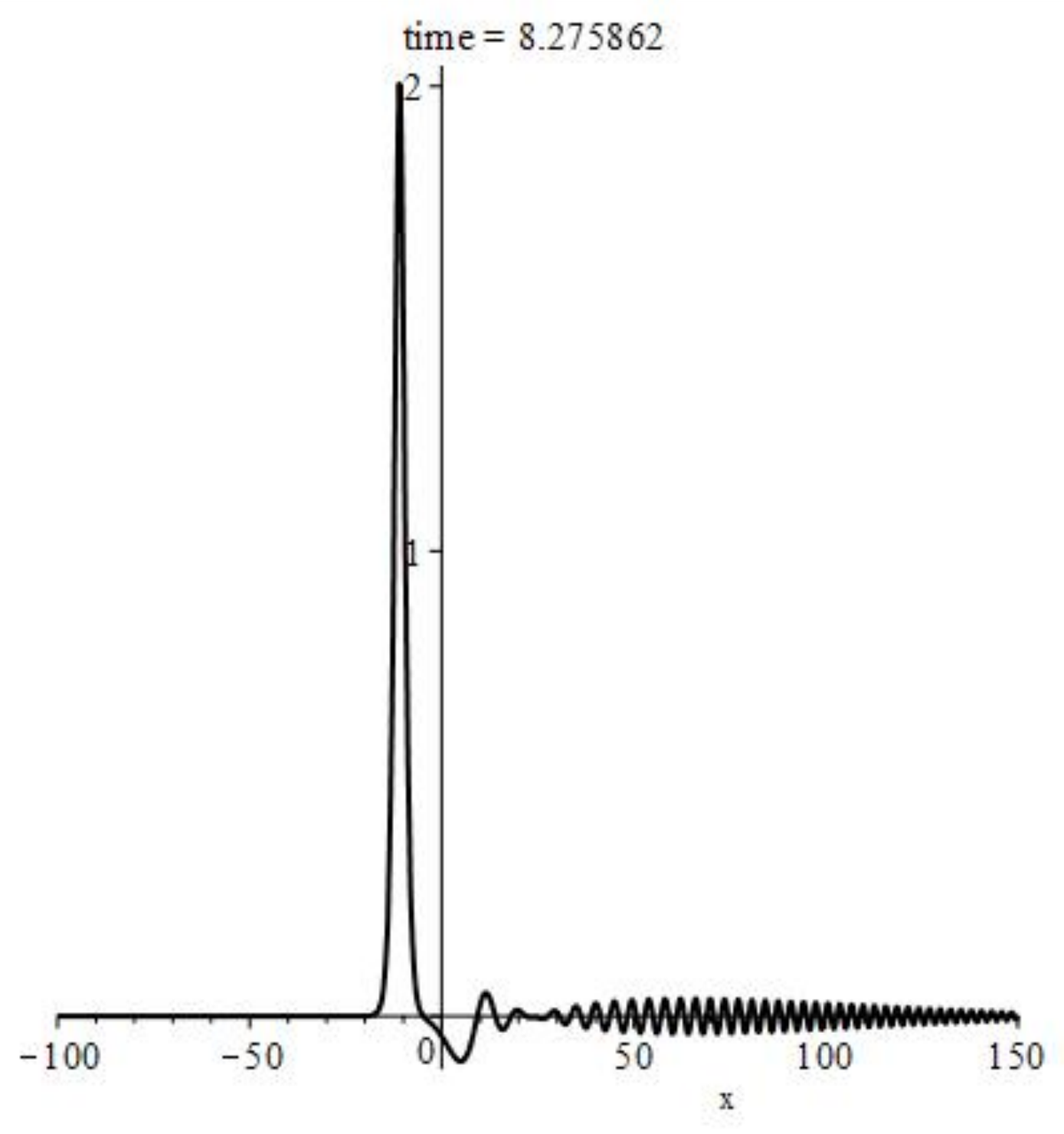}
\end{minipage}
\begin{minipage}{13.2pc}
\includegraphics[width=13.2pc]{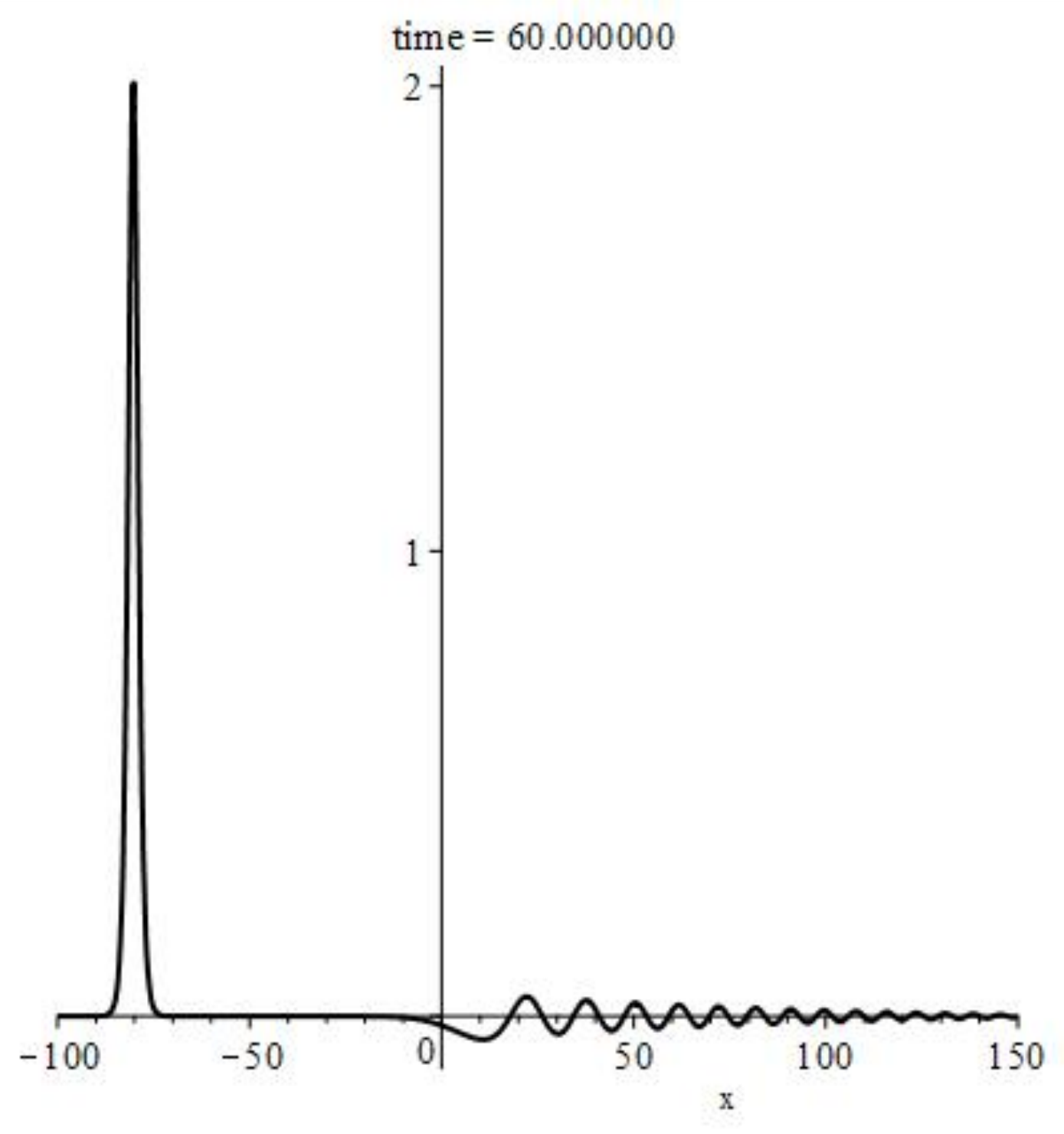}
\end{minipage}
\caption{\textsl{Splitting of the initial perturbation\protect\newline $q(x)=1+ \cos(x), \mbox{ on } x\in [-\pi,\pi]$}, 
\textsl{\textbf{Left}: $t=8$. \textbf{Right}:  $t=60$.}}
\label{8}
\end{figure}

The amplitude of the resulting soliton can be measured wih high precision. It is $2.005=6a_1^2$ , see figure \ref{8}, so $a_1\approx 0.578$. The inequalities

\[I_1(q)=2\pi\leqslant 12a_1,\;I_2(q)=3\pi\geqslant 48a_1^3,\;I_3(q)=7\pi\leqslant \frac{1728}{5}a_1^5\]
hold:
\[\frac{\pi}{6}\approx 0.524<\sqrt[5]{\frac{35\pi}{1728}}\approx 0.576<a_1<\sqrt[3]{\frac{\pi}{16}}\approx 0.581.\]

The system $a_1+a_2=\frac{\pi}{6},];a_1^3+a_2^3=\frac{\pi}{16}$ does  admit non-positive solution $ (0.581,-0.058)$, and $ a_1\approx 0.578, a_2=0$ is an admissible point, nearest to it, see figure \eqref{28}.

Also note that we obtained the conserved quantities for the radiation tail. They are discrepancies $\{I_l(R)=I_l(q)-K_l\sum_{i=1}^n a_i^{2l-1}\}$. In particular, in this example $l=n=1$, $K_1=12$ and the mass of the tail is $I_1(R)=2\pi-12\cdot 0.578=-0.065$

\subsection{\underline{2-soliton}  $q(x)=4(1+ \cos(x)), \mbox{ on } x\in [-\pi,\pi]$\hfill}

\begin{figure}[h]
 \begin{minipage}{13.2pc}
\includegraphics[width=13.2pc]{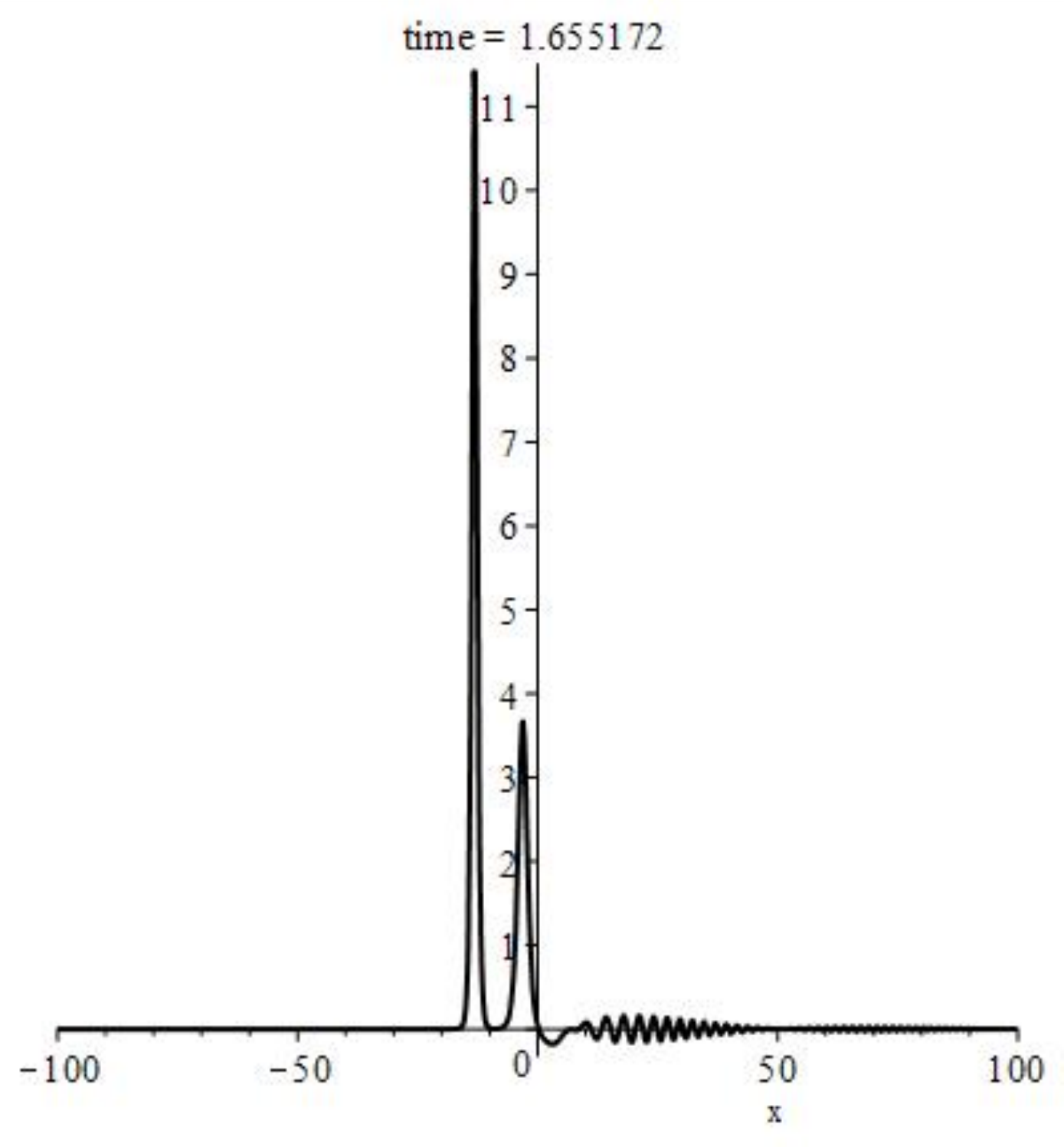}
\end{minipage}
\begin{minipage}{13.2pc}
\includegraphics[width=13.2pc]{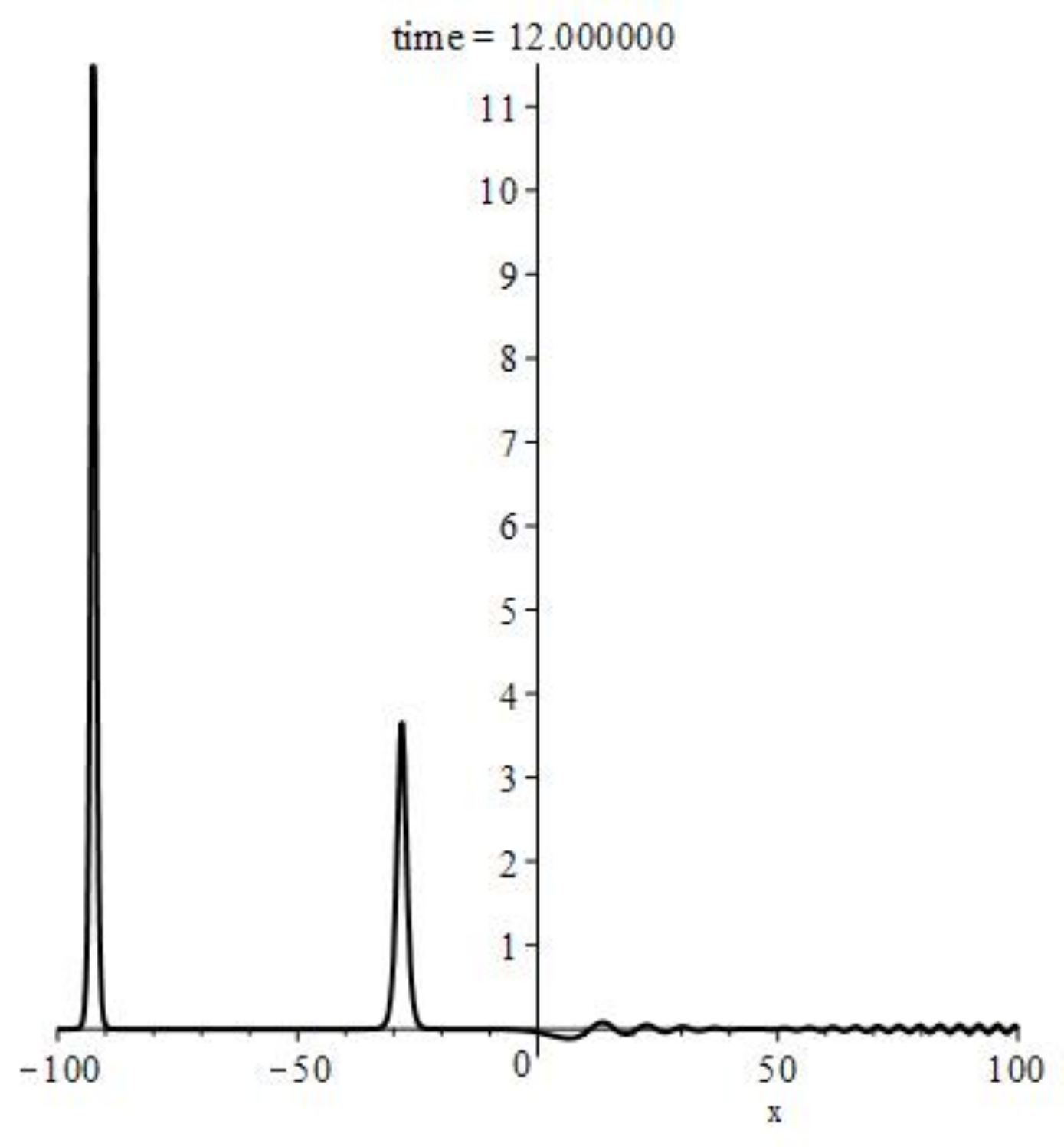}
\end{minipage}
\caption{\textsl{Splitting of the initial perturbation\protect\newline $q(x)=4(1+ \cos(x)), \mbox{ on } x\in [-\pi,\pi]$}, 
\textsl{\textbf{Left}: $t=1.7$. \textbf{Right}:  $t=12$.}}
\label{22}
\end{figure}

In this example $I_1(q)=8\pi,\;I_2(q)=48\pi,\;I_3(q)=592\pi$, see figure \ref{22}.

The number of solitons $n>\sqrt{\frac{p_1^3}{p_2}}=\sqrt{\pi^3/\frac{2\pi}{3}}\thickapprox 1.7\Rightarrow n=2$

The corresponding system for 2-soliton  is $a_1+a_2=\frac{2\pi}{3},];a_1^3+a_2^3=\pi$ has a solution $a_1 = 1.414, a_2 = .681$, while system on 3-solitons has no positive solutions.

The amplitudes of the resulting two solitons measure $11.51=6a_1^2$, $3.65=6a_2^2$, so $a_1\approx 1.385$ and $a_2\approx 0.780 $. The inequalities hold:
   \[\frac{2\pi}{3}\approx 2.094<a_1+a_2\approx 2.165; \]
   \[a_1^3+a_2^3\approx  3.131<\pi\approx 3.142.\]
\[\frac{5\cdot 592\pi}{1728}\approx 5.381<a_1^5+a_2^5\approx  5.383.\]

\subsection{\underline{3-soliton} $q(x)=8(1+ \cos(x)), \mbox{ on } x\in [-\pi,\pi]$}

\begin{figure}[h]
 \begin{minipage}{13.2pc}
\includegraphics[width=13.2pc]{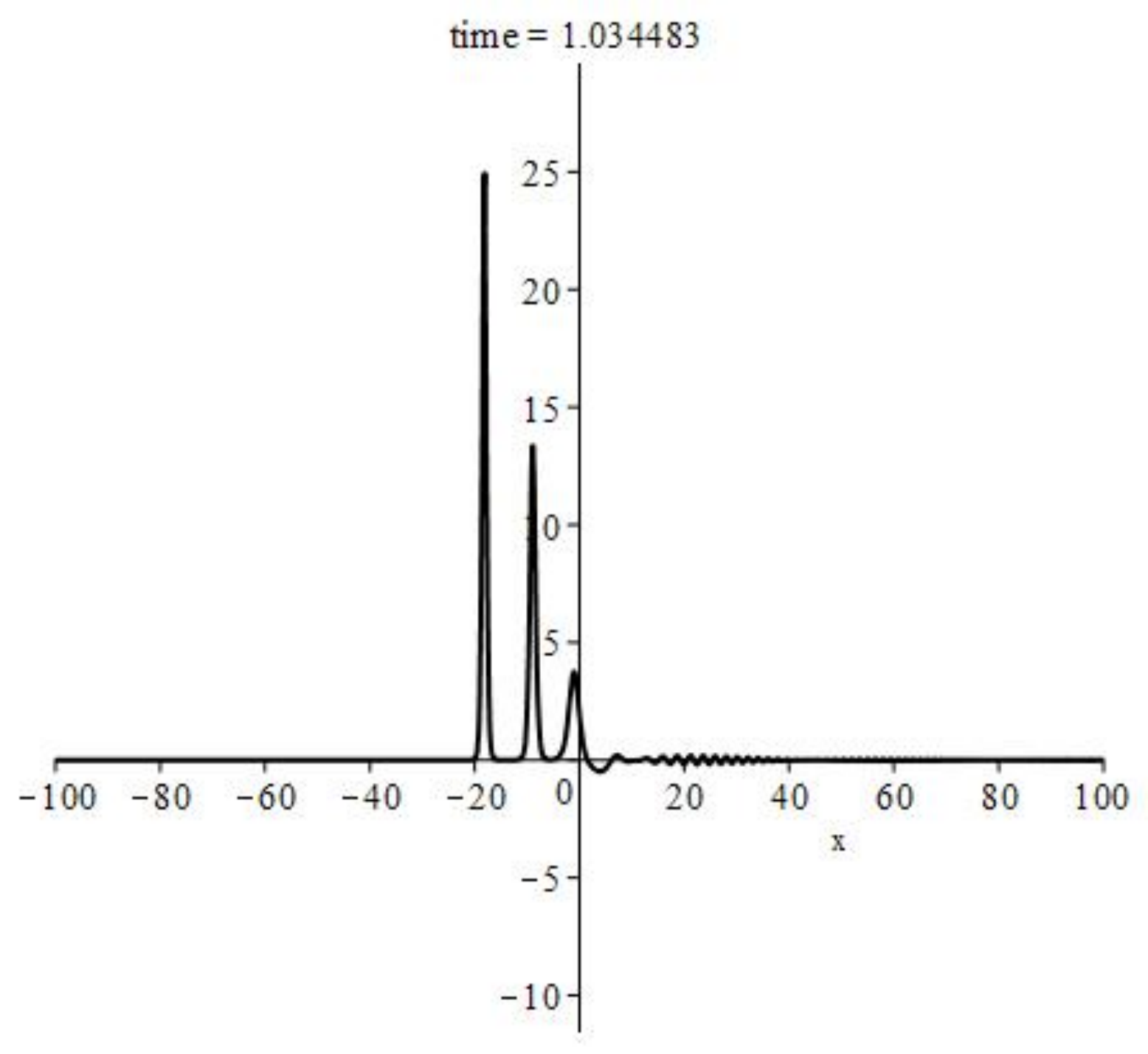}
\end{minipage}
\begin{minipage}{13.2pc}
\includegraphics[width=13.2pc]{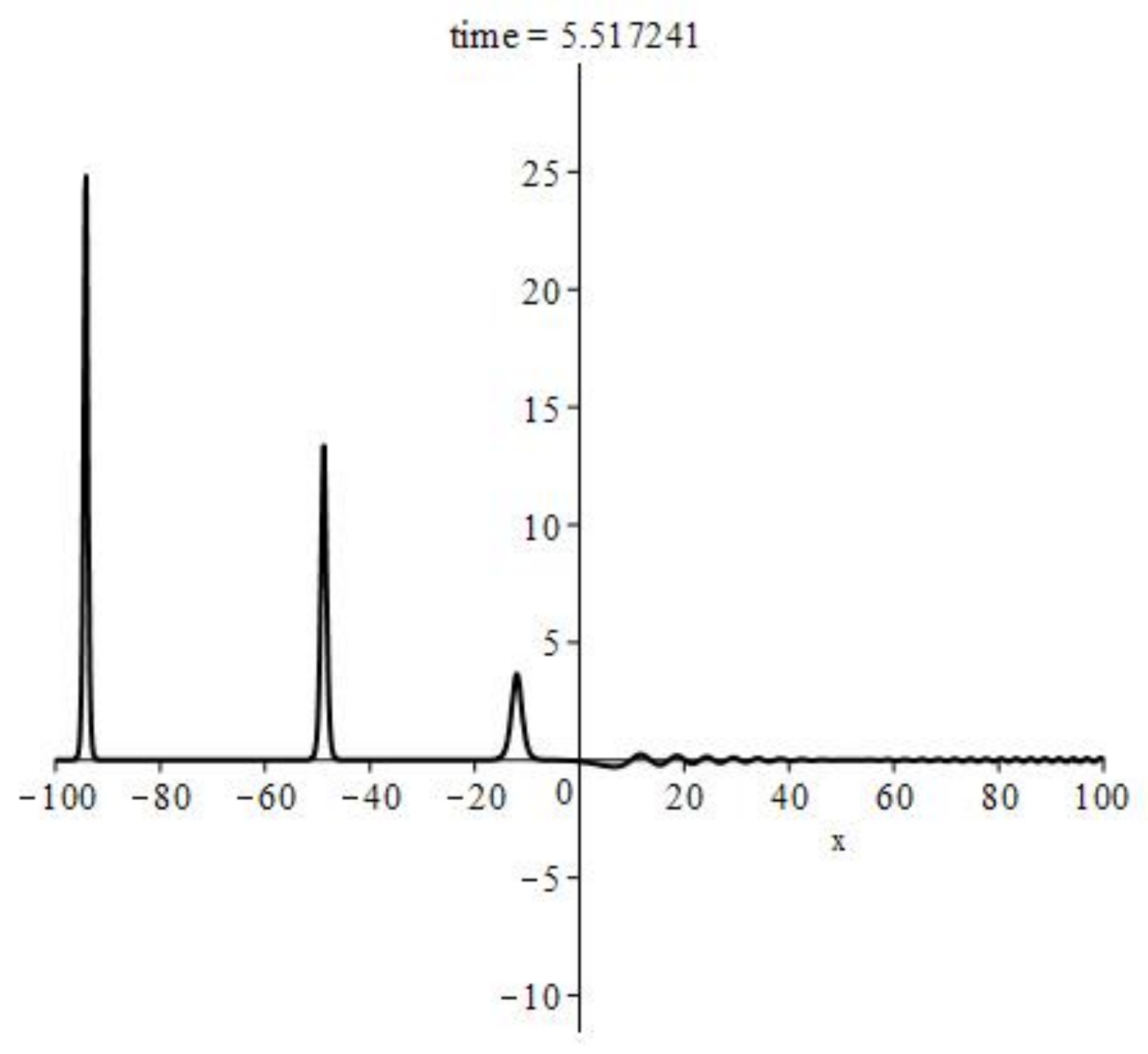}
\end{minipage}
\caption{\textsl{Splitting of the initial perturbation\protect\newline $q(x)=8(1+ \cos(x)), \mbox{ on } x\in [-\pi,\pi]$}, 
\textsl{\textbf{Left}: $t=1$. \textbf{Right}:  $t=5.5$.}}
\label{2}
\end{figure}

In this example $I_1(q)=16\pi,\;I_2(q)=192\pi,\;I_3(q)=4928\pi$, see figure \ref{2}.

The number of solitons $n>\sqrt{\frac{p_1^3}{p_2}}=\sqrt{\frac{4\pi^3}{3}/\frac{4\pi}{3}}\thickapprox 2.4\Rightarrow n=3$

The corresponding system for 3-soliton  is $a_1+a_2+a_3=\frac{4\pi}{3},];a_1^3+a_2^3+a_3^3=4\pi,\;$ $a_1^5+a_2^5+a_3^5=\frac{5\cdot 4928\pi}{1728}$ has a solution $a_1 = 2.034, a_2 = 1.583, a_3 =0.572$.

The amplitudes of the resulting solitons measure $25.5=6a_1^2$, $13.4=6a_2^2$, $3.7=6a_3^2$, so $a_1\approx 2.062 $, $a_2\approx 1.492 $ and $a_3\approx 0.780 $. The inequalities hold:

   \[\frac{4\pi}{3}\approx 4.189<a_1+a_2+a_3\approx  4.333; \]
   \[\frac{5\cdot 4928\pi}{1728}\approx 44.797<a_1^5+a_2^5+a_3^5\approx  44.9100;\]
   \[a_1^3+a_2^3+a_3^3\approx 12.555 <4\pi\approx 12.567 .\]

\subsection{\underline{5-soliton} $q(x)=0.4(-\tanh(x-15)+\tanh(x+15))$}

\begin{figure}[h]
 \begin{minipage}{13.2pc}
\includegraphics[width=13.2pc]{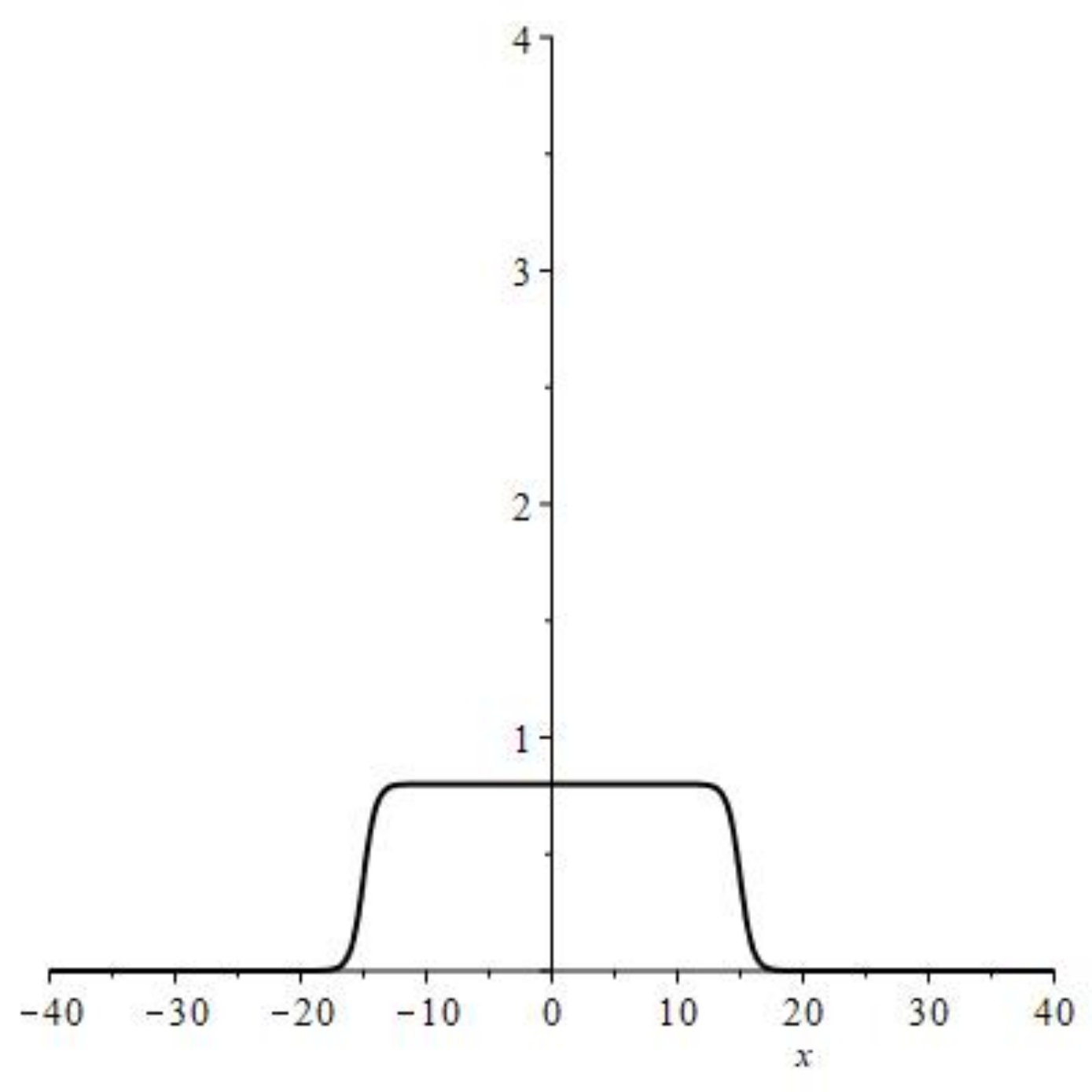}
\end{minipage}
\begin{minipage}{13.2pc}
\includegraphics[width=13.2pc]{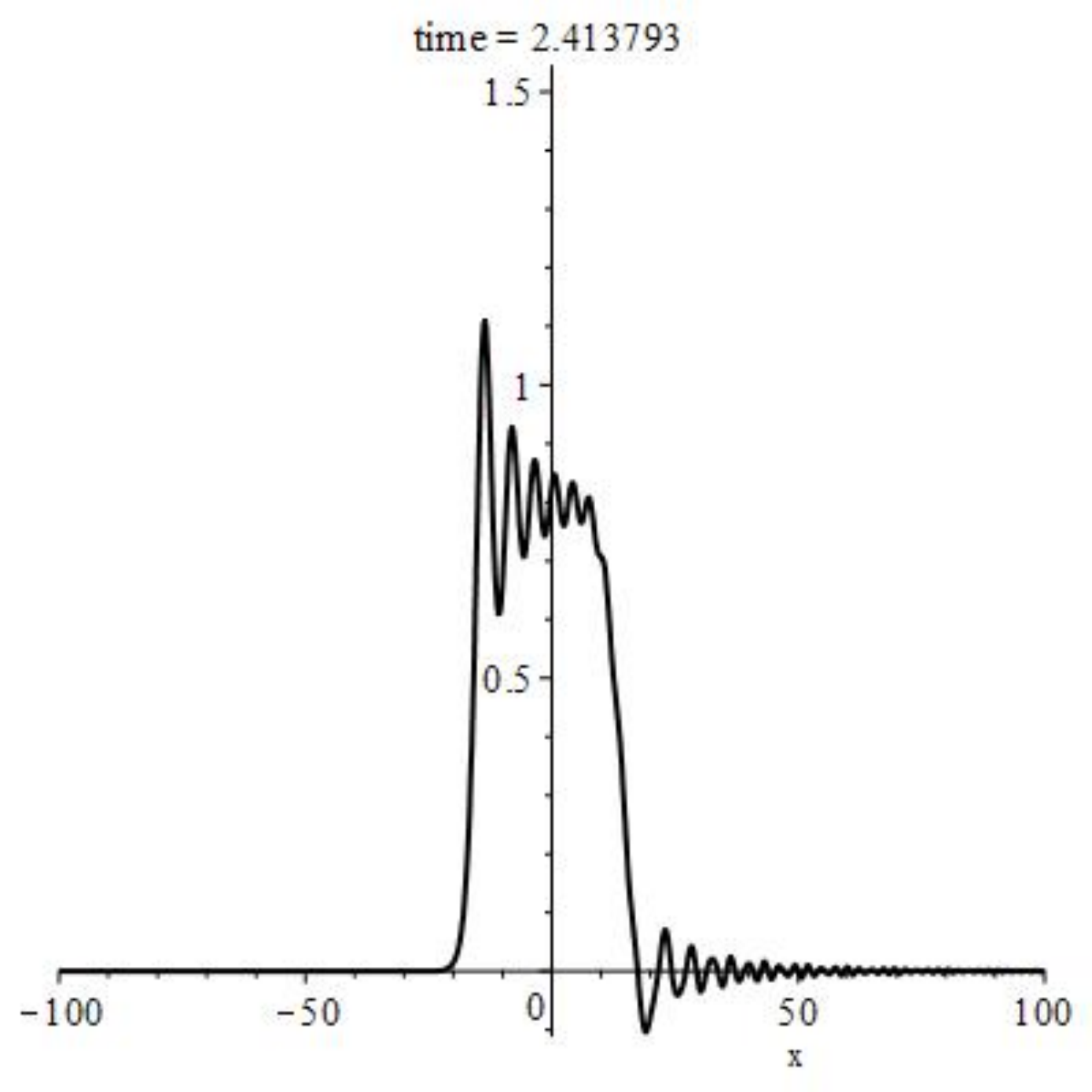}
\end{minipage}
\caption{\textsl{\textbf{Left}: Initial perturbation $q(x)=0.4(-\tanh(x-15)+\tanh(x+15))$.  
\textbf{Right}: Splitting begins as a shock wave,$t=2.4$.}}
\label{43}
\end{figure}

\begin{figure}[h]
 \begin{minipage}{13.2pc}
\includegraphics[width=13.2pc]{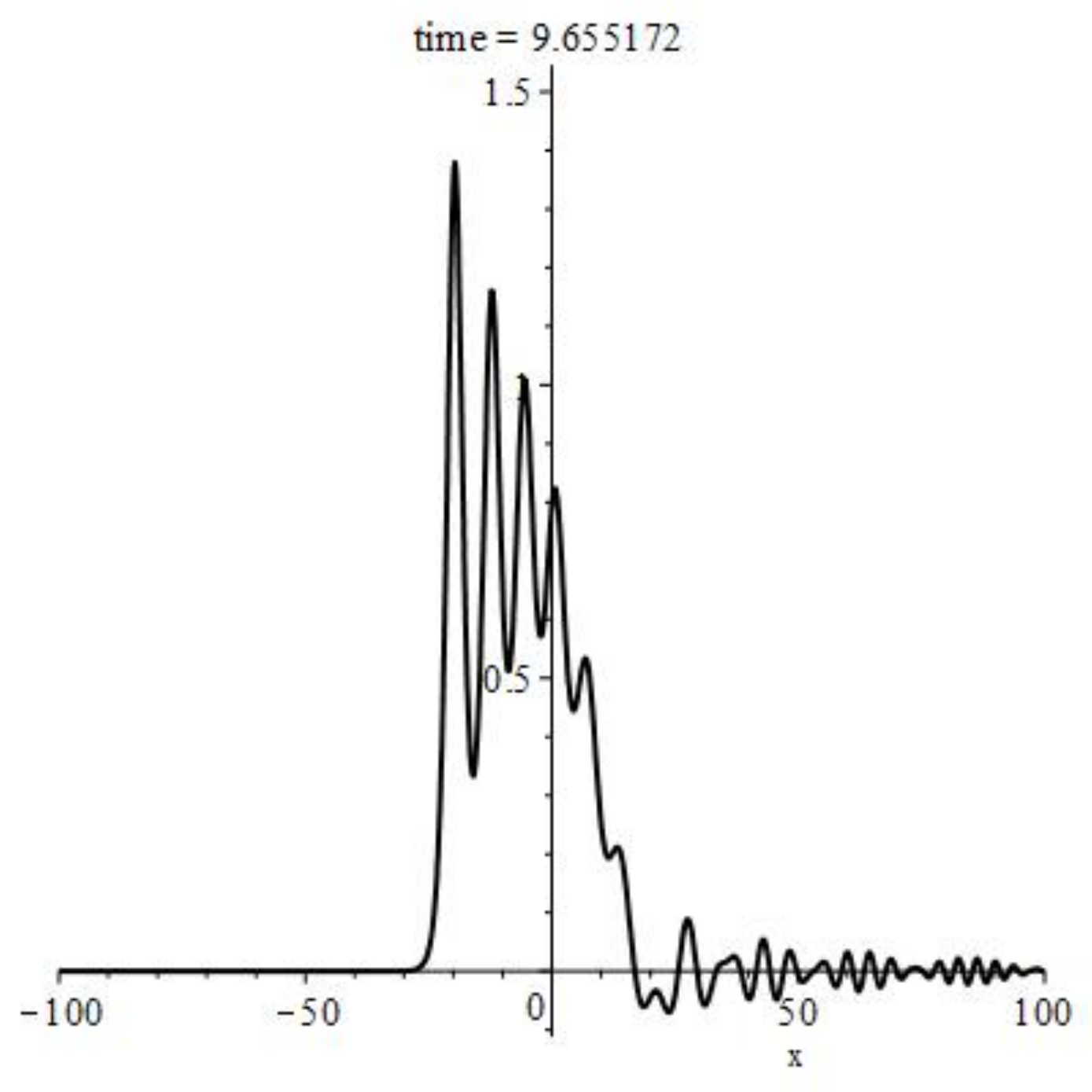}
\end{minipage}
\begin{minipage}{13.2pc}
\includegraphics[width=13.2pc]{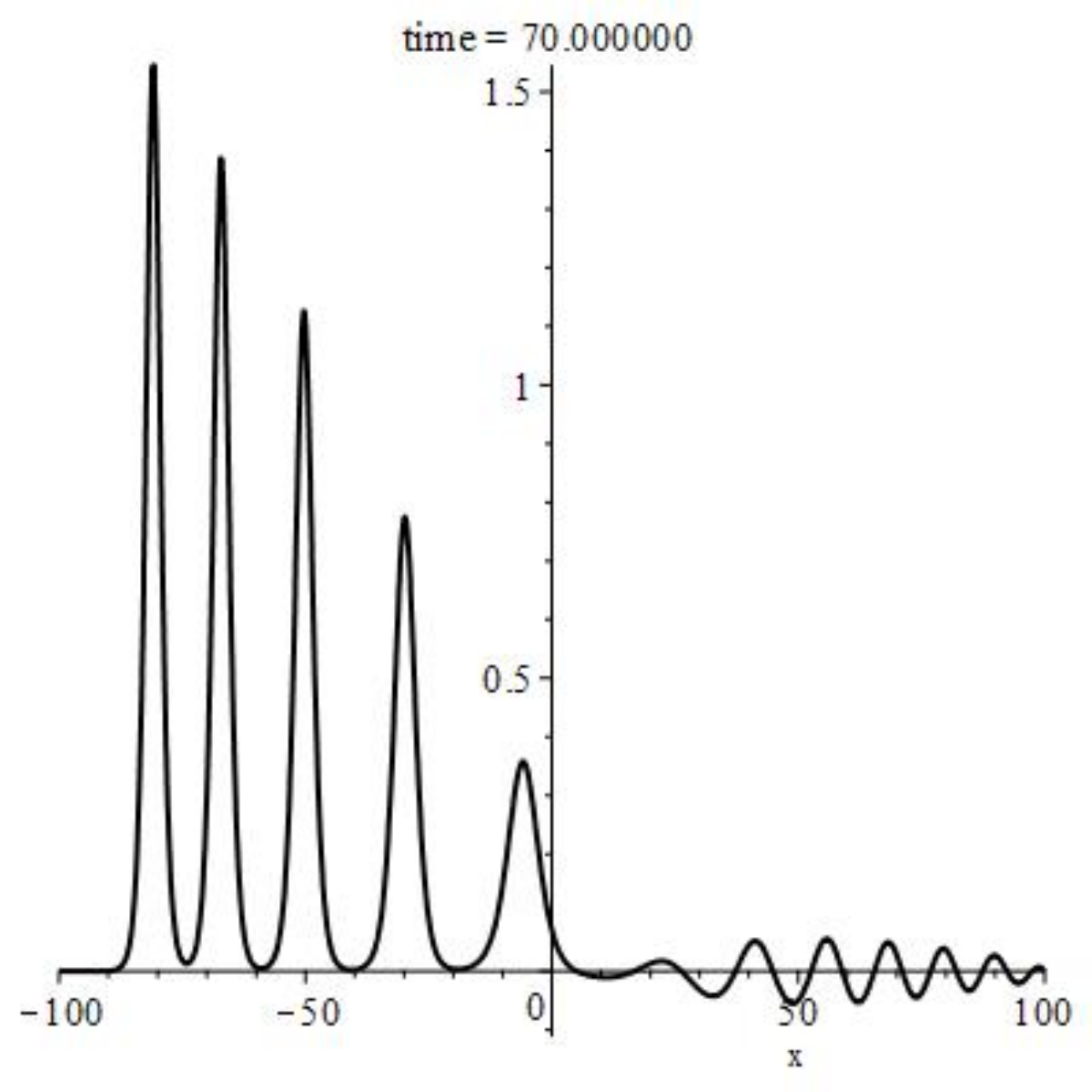}
\end{minipage}
\caption{\textsl{Splitting of the initial perturbation\protect\newline $q(x)=0.4(-\tanh(x-15)+\tanh(x+15))$}, 
\textsl{\textbf{Left}: $t=9.7$. \textbf{Right}:  $t=70$.}}
\label{53}
\end{figure}

In this example $I_1(q)=24\;I_2(q)= 18.56,\;I_3(q)=27.904,\;I_4(q)=55.637$, see figures \ref{43}, \ref{53}

The number of solitons $n>\sqrt{\frac{p_1^3}{p_2}}=\sqrt{\frac{2^3}{0.3867}}\thickapprox 4.5\Rightarrow n=5$

The corresponding system for 4-solitons, $\sum_{i=1}^4a_i^{2j-1},\;j=1,\dots,4$  has no solutions.

The amplitudes of the resulting five solitons measure $1.543=6a_1^2$, $1.385=6a_2^2$, $1.125=6a_3^2$, $0.775=6a_4^2$, $0.36=6a_5^2$, so
                     $a_1\approx 0.507 $, $a_2\approx 0.480$, $a_3\approx 0.433$, $a_4\approx 0.359$, $a_2\approx 0.245$.
The inequalities hold:
\[K_j\sum_{i=1}^5a_i^{2j-1}\geqslant I_j(q)\mbox{ for } j=1,3 \mbox{ and } K_j\sum_{i=1}^5a_i^{2j-1}\leqslant I_j(q)\mbox{ for } j=2,4 \]

\subsection{\underline{2-soliton} $q(x)=1+ \cos(2x), \mbox{ on } x\in [-\pi/2,3\pi/2]$}

\begin{figure}[h]
 \begin{minipage}{13.2pc}
\includegraphics[width=13.2pc]{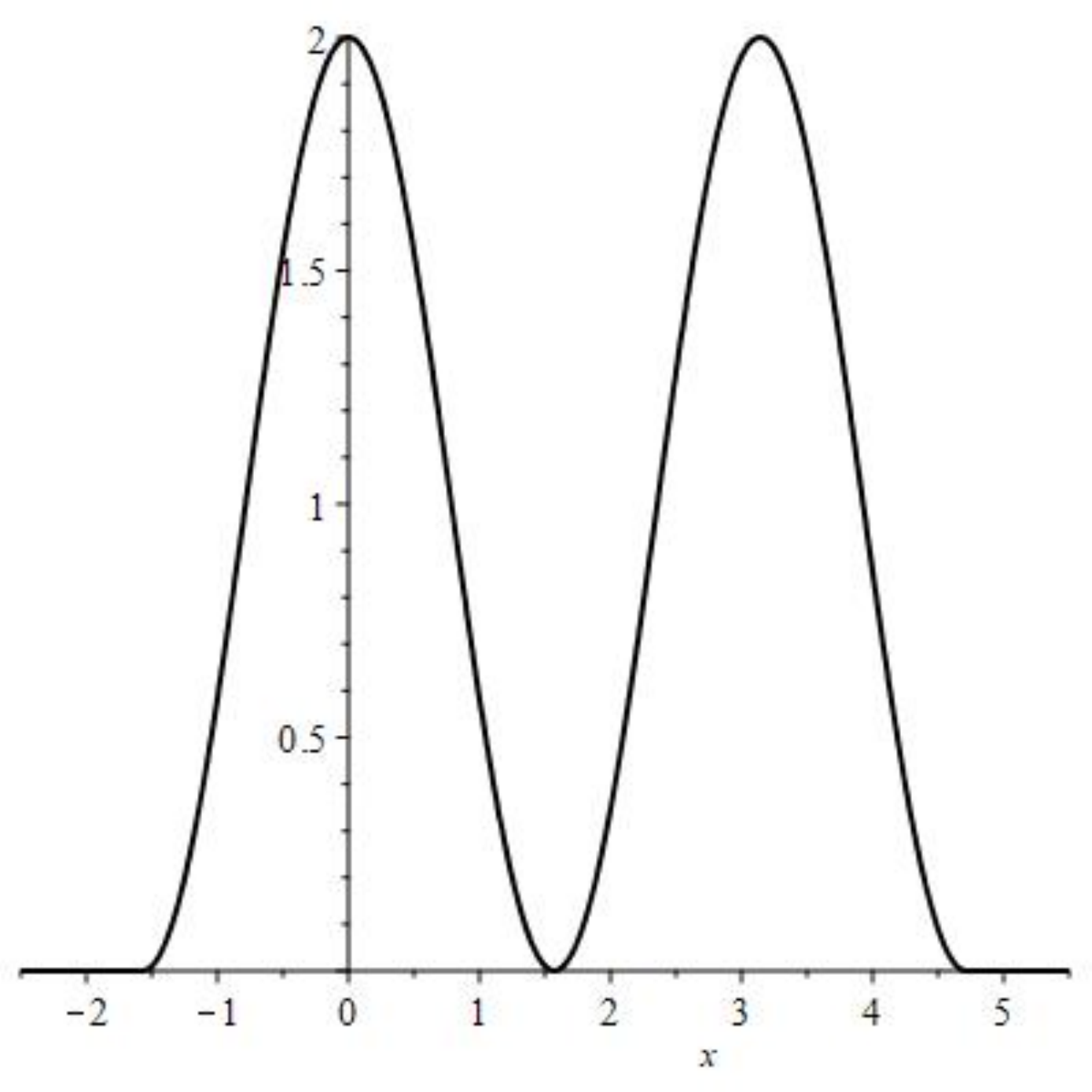}
\end{minipage}
\begin{minipage}{13.2pc}
\includegraphics[width=13.2pc]{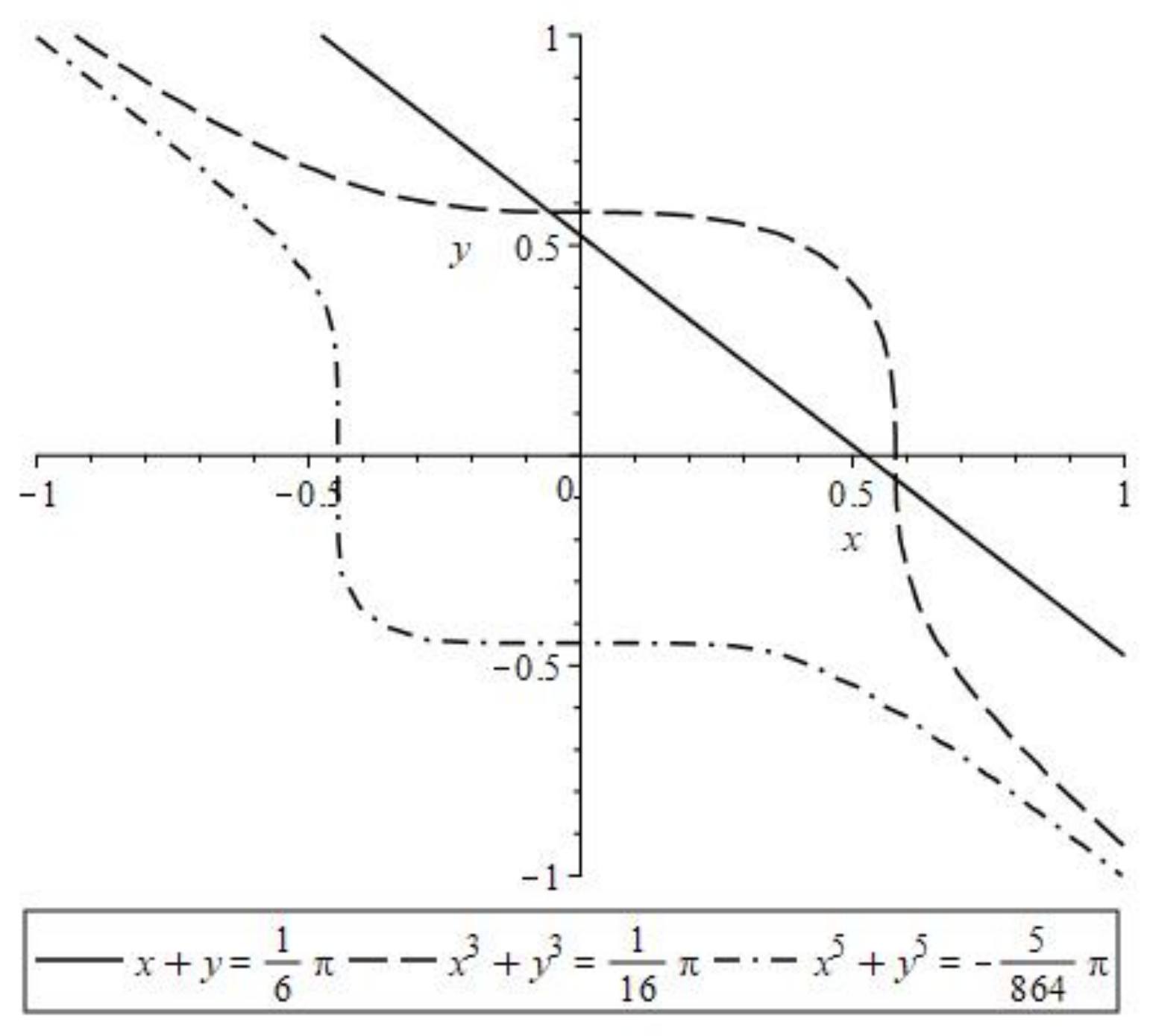}
\end{minipage}
\caption{\textsl{\textbf{Left}: Initial perturbation $q(x)=1+ \cos(2x)$; $t=0$.  
\textbf{Right}: Admissible domain is bounded by dash,  solid lines and $y $ axis.}}
\label{23}
\end{figure}

\begin{figure}[h]
 \begin{minipage}{13.2pc}
\includegraphics[width=13.2pc]{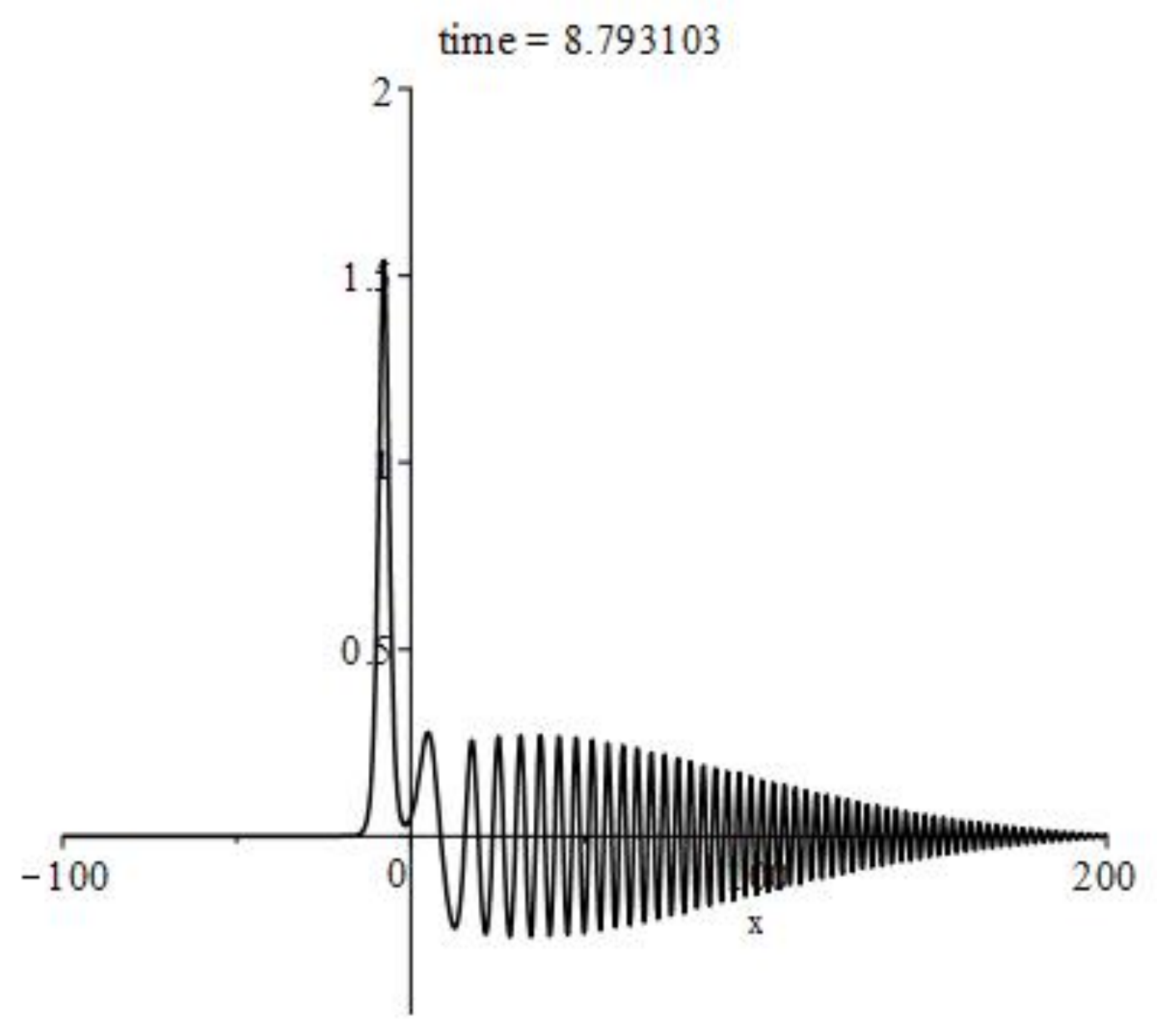}
\end{minipage}
\begin{minipage}{13.2pc}
\includegraphics[width=13.2pc]{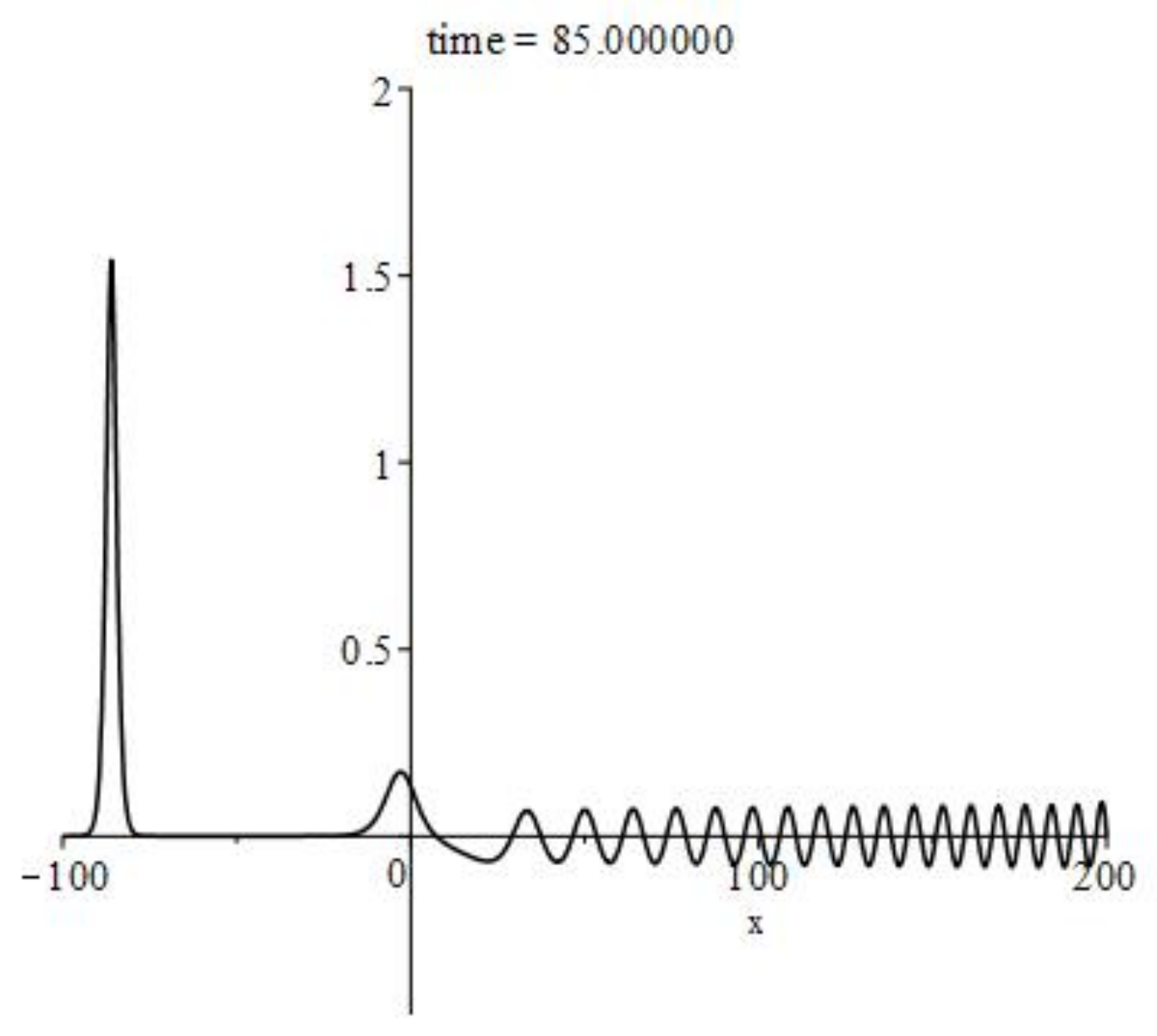}
\end{minipage}
\caption{\textsl{Splitting of the initial perturbation\protect\newline $q(x)=1+ \cos(2x), \mbox{ on } x\in [-\pi/2,3\pi/2]$}, 
\textsl{\textbf{Left}: $t=8.8$. \textbf{Right}:  $t=85$.}}
\label{33}
\end{figure}

In this example $I_1(q)=2\pi,\;I_2(q)=3\pi,\;I_3(q)=-2\pi$, see figure \ref{33}. Note that mass and momentum coincide with those in the first example, so $n>1$. But in contrast to the example 1, $n=2$ since admissible domain is much larger in this case: the third inequality produce no restrictions in positive domain, see figure \ref{23}.

The corresponding system for 3-solitons, $a_1+a_2+a_3=\frac{\pi}{6},\; a_1^3+a_2^3+a_3^3=\frac{\pi}{16},$ $a_1^5+a_2^5+a_3^5=-\frac{10\pi}{1728}$, has no solutions.

The amplitudes of the resulting solitons measure $1.55=6a_1^2$, $0.175=6a_2^2$, so $a_1\approx  0.508  $, $a_2\approx 0.1731 $. The inequalities hold:
 \[\sqrt[5]{\frac{-10\pi}{1728}}<\frac{\pi}{6}<a_1+a_2\]
 \[a_1^3+a_2^3\thickapprox 0.136<\frac{\pi}{16}\]

 The 2-soliton system $a_1+a_2=\frac{\pi}{6},\; a_1^3+a_2^3=\frac{\pi}{16}$ does  admit non-positive solution $ (0.581,-0.058)$, and $a_1, a_2$ is the admissible point, not far from this solution, see figure \ref{23}.

\textbf{Remark.} It must be noted here that the described method is not as  effective when the initial data consists of a disjoint union of perturbations. Later generated solitons in this case collide with tails of previous solitons and a whole picture becomes tangled, at least for a initial short period.

\section*{Conclusion}
The present paper as well as our previous research of the KdV solitons in nonhomogeneous media (\cite{key-5}--\cite{key-10}) persuades that a distorted by inhomogeneity compact impulse getting into homogeneous region behaves  according the same scenario: it became a soliton or splits into two or more. Usually, but not necessarily, the obstacle generates a reflected wave. This effect has the same nature as the splitting  of the initial compact perturbation (or potential) into solitons and a radiation tail in the case of the classical KdV; the number and parameters of resulting solitons vary, but the scenario stays invariable.

We connected the number, amplitudes and velocities of a train  of solitons that is result of a splitting  of an arbitrary initial compact datum for the KdV with its conservation laws; some rough estimations are exemplified above.

 A form of a transformed wave, its reflection and  refraction coefficients may be easily  predicted. Thus the possibility of control of solitary impulses arises. So the results may be of a practical use.

The figures in this paper were generated numerically using Maple PDETools package. The  mode of operation uses the default Euler method, which is a centered implicit scheme, and  can be used to find solutions to PDEs that are first order in time, and arbitrary order in space, with no mixed partial derivatives.

\subsection*{Acknolegement}

This work was partially supported by the Russian Basic Research Foundation grant 18-29-10013.

\end{document}